# Resistivity Minimum in Granular Composites and Thin Metallic Films.

A.Gerber, I.Kishon, D.Bartov and M.Karpovski

School of Physics and Astronomy,

Tel Aviv University

Ramat Aviv, 69978 Tel Aviv, Israel

We analyze the temperature dependence of conductivity in thick granular ferromagnetic compounds $NiSiO_2$ and in thin weakly coupled films of Fe, Ni and Py in vicinity of metal-insulator transition. Development of resistivity minimum followed by a logarithmic variation of conductivity at lower temperatures is attributed to granular structure of compounds and thin films fabricated by conventional deposition techniques. Resistivity minimum is identified as a transition between temperature dependent intra-granular metallic conductance and thermally activated inter-granular tunneling.

## Introduction.

There is a number of universal features demonstrated by variety of materials in vicinity of metal-to-insulator transition. One of them is existence of the resistivity minimum at temperature $T_{min}$, usually followed by a logarithmic increase of resistivity with decreasing temperature. The effect was observed in numerous thin, considered as two-dimensional, and thick, considered as three-dimensional, crystalline and amorphous normal metals [1- 3], paramagnetic and ferromagnetic materials [4 - 8] and superconductors in their normal state [9, 10]. Historically, the phenomenon of resistivity minimum associated with very small amount of paramagnetic impurities in crystalline and amorphous alloys was understood as arising from the spin flip scattering of conduction electrons off the local magnetic moments, randomly distributed in the alloys [11]. Within the accepted Kondo theory, the minimum is suppressed with increasing impurity concentration and finally washed out as magnetic ordering sets in, since magnetic ordering of the local spins destroys the freedom of the spin to flip, the basic mechanism needed to have a resistivity minimum. Thus, observation of the effect in ferromagnetic materials below the magnetic ordering temperature [4 – 8] raised a question of fundamental conceptual importance: since magnetic ordering destroys the Kondo explanation, what then is the source of the observed resistivity minimum? While much work was done in attempts to settle Kondo model with macroscopic ferromagnetism [12, 13] it was noted by Cochrane et al [6] that the logarithmic dependence of resistivity below $T_{min}$ in different ferromagnetic materials fabricated by different methods by different groups was totally unaltered by application of high magnetic field and thus might be not related to magnetism at all.

The next round of interest to the effect came with prediction and observation of quantum corrections to resistivity of disordered materials: weak localization and electron-electron interactions [14, 15]. Both mechanisms predict similar logarithmic temperature dependence of conductivity, but in two-dimensional cases only. Experimentally, not only thin disordered films but also thick three-dimensional granular materials in vicinity of the metal- insulator transition were found to demonstrate the logarithmic temperature variation of resistivity. It was suggested by Deutscher et al. [16] that although thick granular films are ostensibly three-dimensional, they consist of a labyrinth structure in which the metal matrix has a connectivity determined by many tenuous links that may be only a few nm wide, and such a structure is neither one – nor three-dimensional, and its fractal dimensionality close to the percolation threshold may be described as two-dimensional. While questionable for three-dimensional cases, observation of resistivity minimum followed by the logarithmic temperature dependence in thin films is now commonly attributed to the onset of quantum corrections [17-24].

The last, so far, interpretation of the logarithmic temperature dependence of conductivity was suggested by Efetov and Tschersich [25, 26] for granular materials both in two- and three-dimensional structures. To remind, granular materials in this context are defined as metal-insulator composites below geometrical percolation threshold where metal grains with bulk lattice structure are surrounded by thin insulator layer. Granular structures should be distinguished from random composites where both metallic and insulator phases play a symmetric role. A rule has been proposed [27] that the structure of thick composite films, containing several layers of grains, is granular when the insulator is amorphous, and random when it is crystalline. Granular composite films composed of metallic grains coated by an amorphous insulator reach the percolation threshold at a higher value of metal volume content, of the order of 50% or more. The threshold value in 3D random composites is close to 20%, the value predicted for a random continuum. Following Efetov and Tschersich [25, 26] resistivity of granular material depends on intergranular tunneling conductance. One defines the normalized tunneling conductance $g_t = \frac{G_t}{G_0}$, where $G_t$ is the tunneling conductance and $G_0 = \frac{e^2}{\hbar} = 2.43 \times 10^{-4}\ \Omega^{-1} = \frac{1}{4.108\ k\Omega}$ is the quantum conductance. The theory is applicable under the condition: $g_g \gg g_t$, where $g_g = \frac{G_g}{G_0}$ is the normalized intragranular conductance with $G_g$ being the intragranular conductance; and thus assumes that tunneling intergranular conductance dominates the overall transport properties. Depending on strength of intergranular tunneling one distinguishes two cases: weakly insulating and strongly insulating. Weakly insulating regime occurs in granular systems with inter-granular tunneling conductance exceeding the quantum conductance $g_t > 1$ and is characterized by logarithmic dependence of conductivity on temperature:

$$\sigma = \sigma_0(1 + C\ln T) \tag{1}$$

both in 2D and 3D materials, where $\sigma_0$ is the classical global Drude conductivity. The strongly insulating range is ascribed to tunnel junction conductance smaller than the quantum one, $g_t < 1$, and is characterized by an exponential variation of conductivity with temperature as:

$$\sigma = \sigma_0 exp[(-B/T)^n] \quad (2)$$

where $n$ can be different from 1 due to e.g. distribution of grain sizes. It is important to emphasize that the meaning of weakly and strongly insulating granular systems in this context is different from the usually accepted terminology of weak and strong localization. The model is calculated for temperatures high enough to suppress the effects of weak localization, intragranular energy level spacing and Coulomb blockade.

As mentioned above, the transport phenomena described by the model [25, 26] are implemented in regime $g_g \gg g_t$. It is an open question whether the regime with $g_g \leq g_t$ can exist in granular materials. The temperature coefficient of intragranular metallic conductivity is negative, while that of the intergranular tunneling one is positive. Tunneling conductivity of monoatomic insulator layer at elevated temperatures in densely packed granular structures can be quite high, exceeding the intragranular conductivity. One can then envision the state at which $g_t(T)$ intersects with $g_g(T)$ with $g_g(T) > g_t(T)$ at temperatures below the intersection and $g_g(T) < g_t(T)$ above. The intersection would be marked by the local maximum of conductivity or minimum of resistivity.

The problem is intertwined with another one: how to determine whether the system is above or below the geometrical percolation threshold? It is generally accepted that composite material demonstrating negative resistivity temperature coefficient at room temperature is below geometrical percolation threshold. Namely, metallic component does not generate an infinite continuous conducting cluster extending across the entire sample but is interrupted by the insulator partitions. Negative resistivity temperature coefficient is due to thermally assisted tunneling across insulating partitions that have higher resistance than that of metallic clusters. It is equally accepted that composite material demonstrating positive temperature resistivity coefficient at room temperature is above the percolation threshold, and current flows along an infinite metallic path uninterrupted by the insulator phase. However, if thickness of the insulator coating is of order of a monolayer its resistance at elevated temperature can be lower than that of metallic cluster it embeds. In this case, the total observable resistance of a sample below percolation threshold would be that of the metallic clusters, despite their finite size. The question then can a composite material demonstrating positive resistivity temperature coefficient at room temperature be below geometrical percolation threshold?

This work is an attempt to elucidate the physical meaning of resistivity minimum in metal-insulator mixtures and in thin weakly coupled films, composed of single metal with no artificial addition of insulating material, fabricated by conventional deposition techniques. Ferromagnetic metals were chosen to avoid the Kondo-type phenomena. We find that three-dimensional Ni-$SiO_2$ mixtures and thin weakly coupled ferromagnetic films of iron, nickel and permalloy demonstrating resistivity minimum are well described by the model of Efetov and Tschersich below $T_{min}$ and can therefore be considered topologically granular below geometrical percolation threshold. We identify the resistivity minimum as transition between regimes dominated by temperature dependent intra-granular metallic conductance above $T_{min}$ and thermally activated inter-granular tunneling below $T_{min}$ .

## Results and discussion.

### NiSiO$_2$

We start with $Ni_x(SiO_2)_{1-x}$ mixtures fabricated by e-beam co-deposition of Ni and $SiO_2$ from two separate crucibles. Fig.1 presents TEM images of two 10 nm thick samples with Ni volume concentrations x = 0.5 (a), and 0.8 (b). Ni and $SiO_2$ are mutually immiscible and form a typical granular structure with crystalline Ni and amorphous $SiO_2$ totally or partially encapsulating Ni clusters. Ni clusters are polycrystalline composed of 3-5 nm Ni crystallites. $Ni_{0.5}(SiO_2)_{0.5}$ sample (Fig.1a) is clearly below Ni continuity threshold with metallic clusters 5-10 nm diameter embedded within thick amorphous $SiO_2$. Metallic clusters grow with Ni concentration while the volume occupied by $SiO_2$ shrinks to narrow finite channels. Visual examination of Fig.1b does not allow to judge whether Ni forms continuous infinite cluster because TEM resolution is not sufficient to identify monolayer-thick $SiO_2$ partitions and estimate their continuity.

We turn now to transport measurements. Films studied here were 200 nm thick, and can be considered three-dimensional [28]. Fig.2 is a general presentation of $Ni_x(SiO_2)_{1-x}$ resistivity as a function of temperature for samples with several Ni volume concentrations. Resistivity of samples is normalized to their room temperature value. The resistance behavior changes gradually from a purely metallic (resistivity decreases with decreasing temperature and saturates to the remnant value in the low temperature limit) in samples with high Ni concentration (x=0.72), to the insulating one in samples with reduced Ni content showing a negative resistivity temperature coefficient $\alpha = \frac{1}{\rho}\frac{d\rho}{dT} < 0$ in the entire temperature range between 1.5K to room temperature (x=0.22). Intermediate samples have resistivity minimum at temperature $T_{min}$ with $\alpha < 0$ at temperatures below $T_{min}$ and $\alpha > 0$ above $T_{min}$. The feature is illustrated in Fig.3 for a number of samples with Ni volume concentrations between x=0.3 to x=0.55. Here, the resistivity is normalized by its value at $T_{min}$. Below $T_{min}$ resistivity varies as logarithm of temperature. Appearance of resistivity minimum in $Ni_x(SiO_2)_{1-x}$ does not depend on the fabrication method; it was also observed in $Ni_x(SiO_2)_{1-x}$ films prepared by RF magnetron sputtering [29].

Following Refs.25, 26 conductivity of granular material below percolation threshold can be described either by Eq.1 in the weakly insulating regime or by Eq.2 in the strongly insulating regime. Fig.4 illustrates a gradual change of the conductivity temperature dependence with reducing metal content. The same data are analyzed either as conductivity versus logarithm of temperature (Eq.1) – left column; or as logarithm of conductivity versus $T^{-n}$ with $n = 1/2$ or $n = 1/4$ (Eq.2) – right column. Samples demonstrating resistivity minimum (conductivity maximum) – Fig.4a, follow the logarithmic temperature dependence at temperatures below $T_{min}$, and as such can be classified as belonging to the weakly insulating granular range. On the other hand, high resistivity samples with low Ni content, demonstrating monotonic decrease of conductivity with decreasing temperature, are well presented by the exponential dependence (see Fig. 4f with $n = 1/4$) and can be classified as strongly insulating. Exact determination of the power index n, 1/2 or 1/4, is rather ambivalent since changes in conductivity are not sufficiently large. Intermediate samples, showing transition between logarithmic to exponential thermal dependence (Figs. 4b,e), correspond to the transition range between the weakly and strongly insulating regimes. Following the terminilogy of Refs. 25, 26 samples can be characterized by their normalized conductance, estimated as: $g = (\rho/a)^{-1}/G_0$, where $\rho$ is resistivity and $a$ an average metallic cluster size. For $a = 10nm$ the calculated values are $g \approx 20$ for the sample with concentration x=0.55 (a,b), demonstrating resistivity minimum and logarithmic conductance; $g \approx 10^{-3}$ for the strongly insulating sample x=0.22 (e,f) showing exponential conductance; and $g \approx 0.3$ for the intermediate sample x=0.26 (c,d). This estimation is in a very good agreement with the theory predicting transition between the logarithmic and exponential behaviors at $g = 1$.

Up to not very high temperatures resistivity of samples, demonstrating resistivity minimum and logarithmic variation of conductance below $T_{min}$, can be presented as:

$$\rho(T) = \rho_0 + aT^{\nu} - A\rho_0^2 lnT \quad (3)$$

where $\rho_0$ is the remnant zero temperature resistivity, $aT^{\nu}$ is the major thermally dependent scattering term, and the last logarithmic term is obtained from Eq.1 assuming that conductivity does not change significantly at low temperatures: $(\sigma_0 - \sigma)/\sigma_0 \ll 1$. Upon minimization, we obtain:

$$T_{min} = \left(\frac{A}{a\nu}\right)^{\frac{1}{\nu}} (\rho_0)^{\frac{2}{\nu}} \quad (4)$$

Figure 5 presents $T_{min}$ as a function of $\rho_{min}$ (we assume $\rho_{min} \approx \rho_0$). The dependence is linear, which gives the power index $\nu = 2$. This value is consistent with classical electron-electron scattering usually found in bulk ferromagnetic materials [30].

Thus, resistivity of samples demonstrating resistivity minimum can be understood as a superposition of temperature dependent intragranular resistivity, typical for ferromagnetic materials, and thermally assisted intergranular tunneling, while $T_{min}$ is temperature at which a positive intragranular resistivity temperature coefficient equals a negative intergranular tunneling one. The fact that behavior of $NiSiO_2$ composites follows so well predictions of the granular model implies that their structure meets the requirements of the model: the material is composed of metallic grains or polycrystalline clusters separated by insulating coating and is below geometrical percolation threshold.

**Thin films.**

The physical meaning of resistivity minimum can also be asked for thin metallic films grown by conventional deposition techniques. Multiple deposited materials form well separated islands at early fabrication stages. With adding material islands grow, cover gradually the plane, fill the voids and coalesce into continuous media. Gaps unfilled by material play the same role as insulator in granular metal-insulator mixtures. It is well known that morphology, crystallite size and thickness at which conductance percolation is reached depend on material, substrate, temperature, deposition rate etc. Typical morphology of thin Fe films is illustrated in Fig. 6 for films 5 nm thick (a) and 7 nm thick (b). Average thickness here is defined as a total mass deposited per area unit divided by bulk density. Dark areas are crystalline Fe, light are unfilled voids. Individual separated metallic grains can be well identified in Fig. 6a. Metallic clusters grow in thicker films. However, unfilled gaps are clearly present also in the 7 nm thick film, and we might wonder whether these voids still form an infinite cluster.

We switch to the transport properties of thin Ni, Fe and Permalloy ferromagnetic films. Data shown here are representative for multiple samples deposited by e-beam deposition or RF sputtering on glass, GaAs or other substrates, patterned to Hall bar geometry or having rectangular form and measured using Van der Pauw protocol. In the following we shall discuss series of Ni and Fe films with various thicknesses and Py films of a constant thickness where variation of resistivity was achieved by post-fabrication annealing. Ni and Fe series with thickness between 2.5 nm and 100 nm were deposited on room temperature substrates. A batch of 17 nm thick Py samples were deposited on sapphire substrates and passed a sequence of mild annealing treatments at temperatures from 100˚C to 250˚C for periods of few hours. Each annealing stage gradually reduced resistivity of a sample, finally reaching about 50% of the initial pre-annealing value.

Fig. 7 presents the temperature dependence of resistivity of a series of Fe films between 10 and 2.5 nm thick. Similar to $NiSiO_2$ composites, resistivity of single-element thin films varies from metallic ($\alpha \geq 0$ between room temperature and 1.5 K) in thick films to insulating ($\alpha < 0$ in the entire temperature range) in thin ones, while films of intermediate thicknesses demonstrate a transition between $\alpha < 0$ and $\alpha > 0$ with resistivity minimum at certain temperature $T_{min}$, as shown in Fig.7 inset for a series of Fe films. In Ni and Fe series we observed $T_{min}$ reaching 90 - 100 K when thickness was reduced from 10 to 3 nm. In Py series $T_{min}$ decreases with annealing from 60K to 30 K. Below $T_{min}$ conductivity of all series follows logarithmic temperature dependence $\Delta\sigma \propto \ln T$, illustrated in Fig.8 for Ni and Py series.

Traditionally, logarithmic dependence of conductance in thin films is attributed to two quantum correction mechanisms: weak localization (WE) and electron-electron interactions (EE). In 2D both mechanisms predict similar temperature dependent corrections to conductance:

$$\Delta\sigma_{2D}^{WE}(T) = \alpha lnT \qquad (5)$$

for weak localization, and

$$\Delta\sigma_{2D}^{EE}(T) = \alpha\left(1 - \tfrac{3}{4}\widetilde{F_\sigma}\right)lnT \qquad (6)$$

for e-e interaction, where $\alpha = e^2/2\pi^2\hbar \cong 1.23 \cdot 10^{-5}\ \Omega^{-1}$, and $\widetilde{F_\sigma}$ is the electron screening factor, defined to range from 0 (no screening) to 1 (complete screening). While the temperature dependence of the two mechanisms is similar, their response to magnetic field is different: weak localization is expected to be suppressed by quite weak fields that induce dephasing of electronic scattering loops. We show in Fig.9 resistance of Fe (main panel) and Py (inset) samples measured as a function of temperature at several fields applied normal to the film plane. 16 Tesla field is a few orders of magnitude above the value sufficient to destroy the coherent effect of weak localization and cause the logarithmic correction to vanish. Application of high fields does not affect $T_{min}$. We therefore dismiss the weak localization as the mechanism responsible for resistivity minimum in ferromagnetic films in agreement with previous studies [17-19, 31]. Vertical shift between the zero and high field curves is due to a combination of magnetoresistance mechanisms: anisotropic magnetoresistance, linear positive magnetoresistance and temperature dependent spin-magnon scattering [31].

Coefficient of the logarithmic conductance dependence is analyzed in Fig.10. Here, the normalized logarithmic slope of conductance $\frac{1}{\alpha}\frac{d\sigma}{dlnT}$ (left axis) and the screening factor $\widetilde{F_\sigma}$ (right axis) are shown as a function of sheet resistance for Fe samples 2.5 -10 nm thick. The screening parameter is extracted as: $\widetilde{F_\sigma} = \frac{4}{3}(1 - \frac{1}{\alpha}\frac{d\sigma}{dlnT})$. Roughly, the measured logarithmic slopes are close to the theoretical $\alpha$ value. However, we observe a clear tendency of growing slope in thinner films with higher sheet resistance. In thin films with resistance above 1000 Ω/□ the measured slope exceeds the maximum expected for 2D electron-electron corrections and the calculated screening factor becomes negative, which is unphysical in the framework of the conventional model. Ni and Py series show a similar trend. Slopes larger than predicted for 2D electron-electron interaction were found in narrow Co nanowires [17, 18] when their width was reduced below few hundred nm. The effect was attributed to reduced dimensionality of the system and transition from 2D to 1D regimes. Such explanation can be hardly accepted for our macroscopically wide films.

We analyzed our data in the framework of the granular model. Following Ref. 32, the slope of the logarithmic temperature dependence of conductivity $C$ (Eq.1) for an arbitrary periodic lattice is given by $C = (\pi z g)^{-1}$, where $g$ is the normalized conductance and $z$ is an average coordination number (an average number of adjacent grains). For the simple case of periodic cubic lattice $z = 2d$, where $d$ is dimensionality; however, in less ordered structures the effective number of neighbors might be different. Fig.11 presents the normalized logarithmic temperature dependence coefficient $C = \frac{1}{\sigma_0}\frac{d\sigma}{dlnT}$ as a function of sheet resistance $R_{sheet}$ (bottom axis) and the normalized conductance ($g = G/G_0$, where $G = 1/R_{sheet}$) for series of Fe (a) and Ni (b) films of various thicknesses. Symbols indicate the experimental data and lines are calculated for z=6 (solid line) and z=4 (dotted line). Both Ni and Fe series appear to follow well the predictions of the granular model with an average of 6 neighbors per grain.

It is interesting to compare explicitly the resistivity minimum in three-dimensional granular samples and in thin single-element films. Inset of Fig. 12 presents the temperature of resistivity minimum $T_{min}$ as a function of resistivity in thick three-dimensional granular $NiSiO_2$ samples and in thin films of Ni. In both systems $T_{min}$ increases linearly with resistivity, however resistivity of $NiSiO_2$ samples is significantly larger than that of Ni films and no correlation between two systems is visible. The main panel of Fig.12 presents the same data as a function of the normalized conductance $g$. For thin Ni films $g$ was calculated as: $g = (R_{sheet})^{-1}/G_0$, and for thick Ni-$SiO_2$ as: $g = (\rho/a)^{-1}/G_0$, where $\rho$ is resistivity and $a$ an average size of Ni clusters, taken as $a = 10nm$. Estimation of parameter $g$ taken here is very rough, nevertheless similarity in behavior of three-dimensional granular Ni-$SiO_2$ and thin weakly coupled Ni films is evident, and one can suggest a universal scaling of resistivity minimum as a function of the normalized conductance $g$.

In three dimensions the model of quantum e-e corrections in macroscopically homogeneous disordered materials can be distinguished qualitatively from the weakly insulating granular model due to different dependencies of conductivity on temperature: $\Delta\sigma \propto T^{1/2}$ and $\Delta\sigma \propto logT$ respectively. We used this to identify the granular model as the source of resistivity minimum in thick Ni-$SiO_2$ films. In strict two-dimensions, both models predict the same logarithmic behavior. Moreover, at low temperatures the effective electron interaction length grows larger than the grain size and the system becomes effectively homogeneous [25, 26], which is identical to the homogeneous Altshuler-Aronov model [15]. Thus, selection of a proper interpretation in 2D depends on material morphology. We favor the granular interpretation in cases described above due to heterogeneous film structure and a common minimum temperature scaling in thick granular Ni-$SiO_2$, which has been proven to follow the granular model, and thin weakly coupled Ni films.

We discussed two types of materials: metal-insulator composites and thin weakly coupled films. An additional model system for which our arguments can be relevant are partially crystallized amorphous films. Swamy et al [33, 34] reported development of resistivity minimum when amorphous as-deposited CoFeB films were annealed. Amorphous CoFeB has high resistivity and negative resistivity temperature coefficient. Annealled samples are polycrystalline with significantly lower resistivity. Granular interpretation of resistivity minima, as suggested by the authors [34], is applicable in this case assuming that metallic grains stay encapsulated by thin layer of remnant non-crystallized amorphous phase serving as tunneling barriers.

"Granular" interpretation of resistivity minimum is quite significant for understanding real materials in vicinity of metal-insulator transition. Observation of resistivity minimum was adapted as an experimental platform for demonstration of the quantum percolation concept in metal-insulator composites [35-37]. It was assumed that materials demonstrating resistivity minimum are above the classical percolation threshold with interconnected metallic clusters forming infinite conductive path. Transition from the metallic behavior ($\alpha > 0$) at $T > T_{min}$ to the insulating one ($\alpha < 0$) at $T < T_{min}$ was attributed to an increased local quantum interference effect at low temperatures along the infinite metallic percolation path. Quantum percolation threshold was defined as a minimum metallic concentration for which the conductivity temperature coefficient is positive ($\alpha \geq 0$) already at zero temperature. Classical percolation threshold was defined by transition between $\alpha < 0$ and $\alpha > 0$ at $T \rightarrow \infty$, which occurs at metal concentration lower than the quantum percolation. "Granular" interpretation is different: material demonstrating resistivity minimum is below classical percolation threshold, and $T_{min}$ indicates the temperature below which the temperature dependent intergranular tunneling conductance falls below the temperature dependent intragranular one. This interpretation does not identify resistivity minimum as an onset of quantum corrections. It is important to stress that these conclusions relate to heterogeneous granular systems and not to high quality homogeneous 2D electron systems like MOSFETs.

## Summary.

To summarize, we analyzed resistivity temperature dependence of thick polycrystalline metal-insulator mixtures Ni-$SiO_2$, thin films of Fe and Ni with variable thickness, and films of Py at different stages of post-fabrication annealing in vicinity of metal-insulator transition. In all these materials there is a range of samples demonstrating a non-monotonic variation of resistivity as a function of temperature with resistivity minimum at temperature $T_{min}$ followed by a logarithmic dependence at lower temperatures. We attribute the effect to the granular structure of materials. Resistivity minimum at temperature $T_{min}$ is identified as the onset of the tunneling dominated regime, below which the temperature dependent intragranular conductance exceeds the temperature dependent intergranular tunneling conductance, the system enters the weakly insulating regime, and logarithmic conductivity is due to dominance of intergranular tunneling. Positive resistivity temperature coefficient at $T > T_{min}$ indicates the intragranular metallicity when intragranular conductance is smaller than the intergranular tunneling and the intragranular resistivity dominates the total one. Following this interpretation, thin and thick granular films demonstrating resistivity minimum are below the geometrical percolation threshold. $T_{min}$ can be any high and not limited to low temperatures expected for quantum phenomena. Granular interpretation of resistivity minimum is applicable to multiple heterogeneous materials, including granular composites, metals with mechanical defaults and cracks, thin films in vicinity of continuity threshold and semiconductors with non-uniform distribution of dopants.

## References.

.

## Figure captions.

Fig. 1. TEM images of $Ni_x(SiO_2)_{1-x}$ films grown by e-beam co-deposition with Ni concentration x = 0.5 (a), and 0.8 (b). Dark areas are crystalline Ni, light are amorphous $SiO_2$.

Fig.2. Normalized resistivity of 200 nm thick $Ni_x(SiO_2)_{1-x}$ films as a function of temperature for several Ni volume concentrations $x$: 0.22, 0.24, 0.26, 0.48 and 0.72. Resistivity is normalized by its room temperature value.

Fig.3. Resistivity as a function of logarithm of temperature and resistivity minima in $Ni_x(SiO_2)_{1-x}$ samples with Ni concentration x: 0.3, 0.35 and 0.55. The data are normalized by the minimum resistivity.

Fig.4. Conductivity versus logarithm of temperature – left column; and logarithm of conductivity versus $T^{-n}$ with $n = 1/2$ (lower axis) and $n = 1/4$ (upper axis) – right column for $Ni_x(SiO_2)_{1-x}$ samples with x = 0.55 (a,b), x = 0.26 (c,d) and x = 0.22 (e,f). $g$ is the calculated normalized conductivity.

Fig.5. $T_{min}$ as a function of $\rho_{min}$ for a series of $Ni_x(SiO_2)_{1-x}$ samples. Solid line is guide for the eye.

Fig.6. TEM images of 5 nm thick (a) and 7 nm thick (b) Fe films grown by RF sputtering on carbon grid. Light areas are unfilled gaps in-between Fe clusters. Scale ruler is 10 nm in (a) and 20 nm in (b).

Fig.6. Temperature dependence of resistivity of a series of Fe films between 10 and 2.5 nm thick. The data are normalized by room temperature resistivity. Inset: normalized resistivity versus

logarithm of temperature for Fe films 10, 7, 6, 5, 4 and 3 nm thick. Resistivity is normalized by $\rho_{min}$.

Fig.7. Change of conductance as a function of temperature in films demonstrating conductance maximum (resistance minimum): (a) Ni films 8, 7, 5 and 4 nm thick; (b) Py film as deposited and after several post-deposition annealing treatments. $\Delta\sigma = \sigma - \sigma(4.2K)$.

Fig.8. Resistivity of 4 nm thick Fe film (main panel) measured as a function of temperature at zero and 16 T magnetic field applied normal to the film plane. Inset: resistivity of Py film (after the third annealing) measured at 0, 2T, 9T and 16 T fields.

Fig.9. Normalized logarithmic slope of conductance $\frac{1}{\alpha}\frac{d\sigma}{dlnT}$ (left axis) and the screening factor $\widetilde{F_\sigma}$ (right axis) as a function of sheet resistance for Fe films 2.5 -10 nm thick. Solid line is guide for the eye.

Fig.10. Normalized logarithmic temperature coefficient $C = \frac{1}{\sigma_0}\frac{d\sigma}{dlnT}$ as a function of sheet resistance $R_{sheet}$ (bottom axis) and the normalized conductance $g = G/G_0$, where $G = 1/R_{sheet}$, (upper axis) for series of Fe (a) and Ni (b) films of various thicknesses. Symbols indicate the experimental data and lines are calculated for z=6 (solid line) and z=4 (dotted line).

Fig.11. $T_{min}$ as a function of resistivity (inset) and normalized conductance $g$ (main panel) for thick granular $NiSiO_2$ samples (crosses) and thin films of Ni (solid circles).

**Figures.**

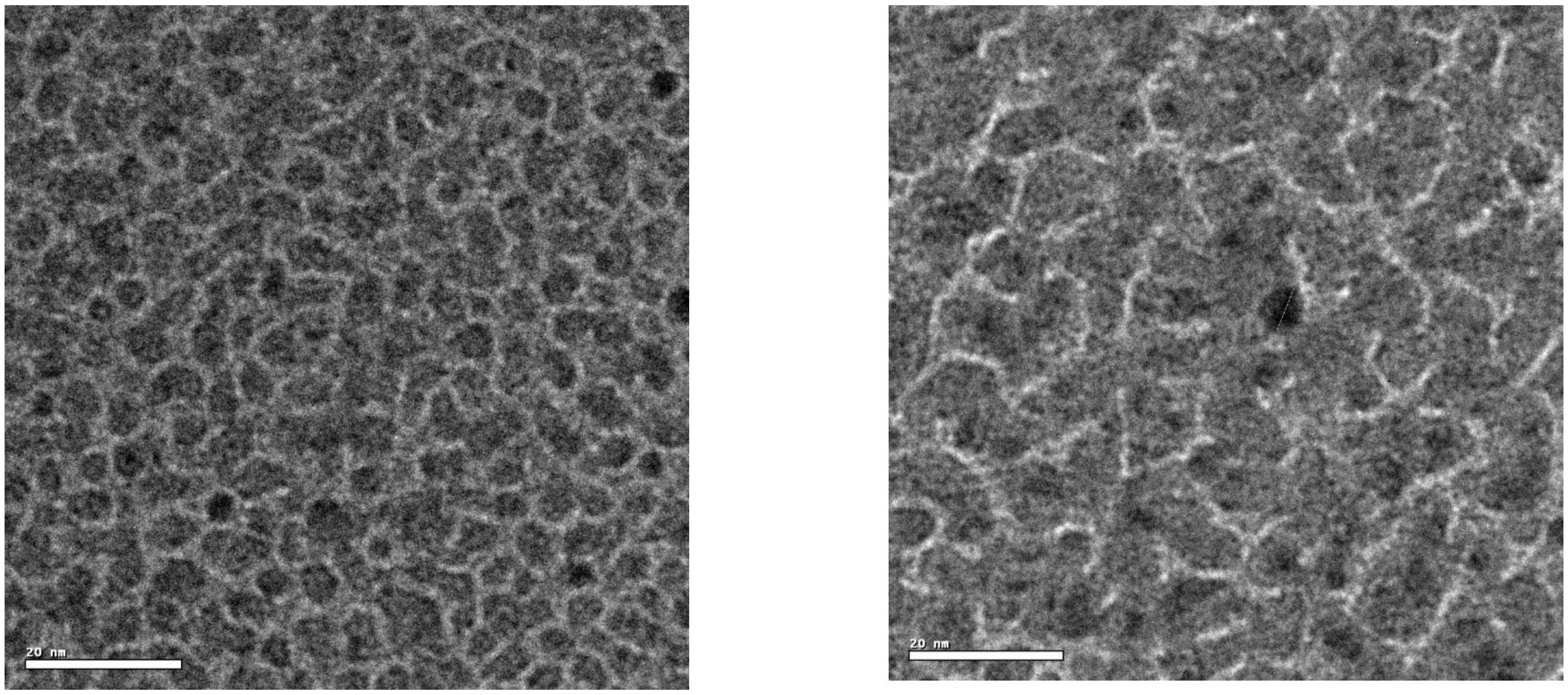

Fig. 1

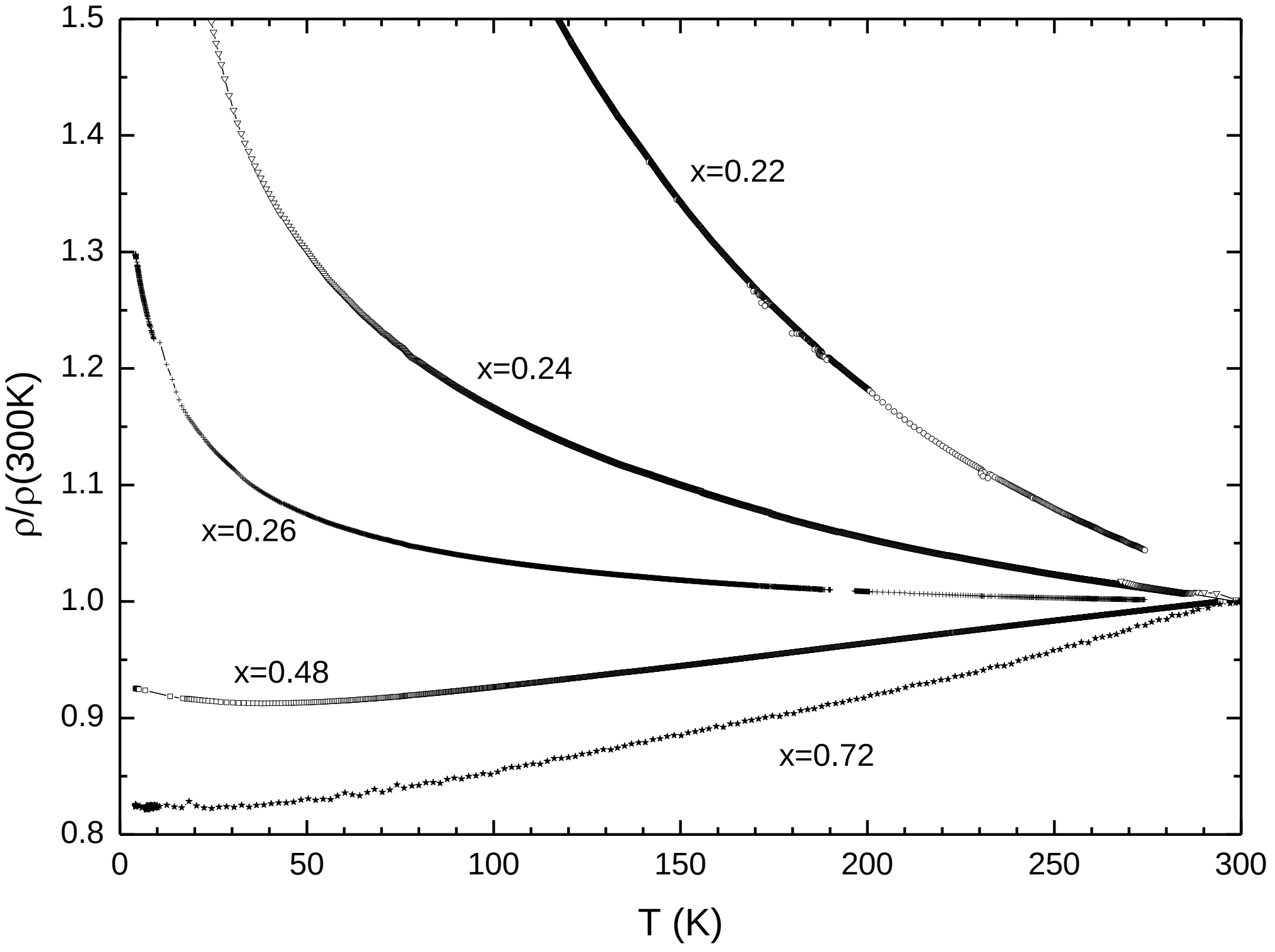

Fig. 2

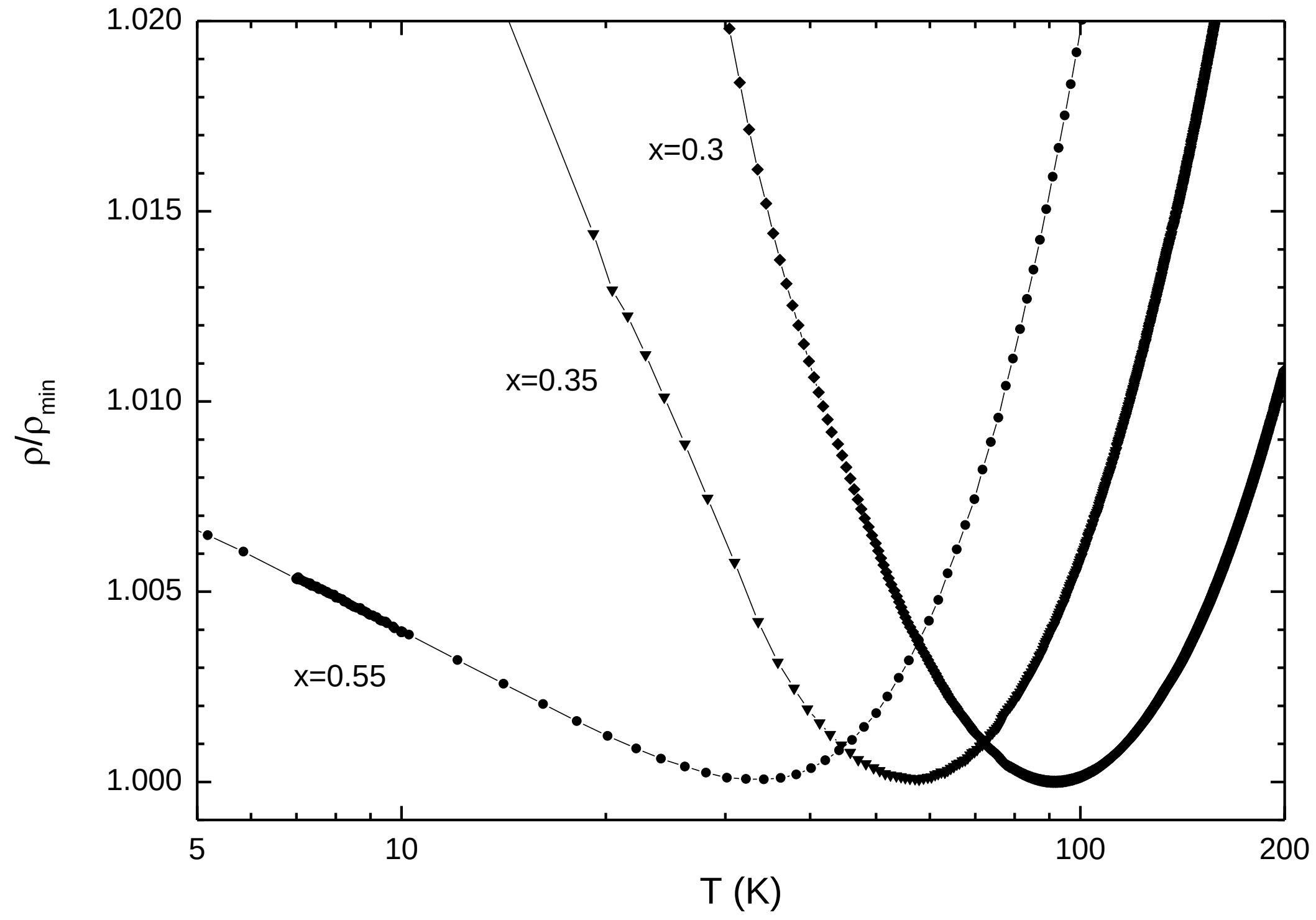

Fig. 3

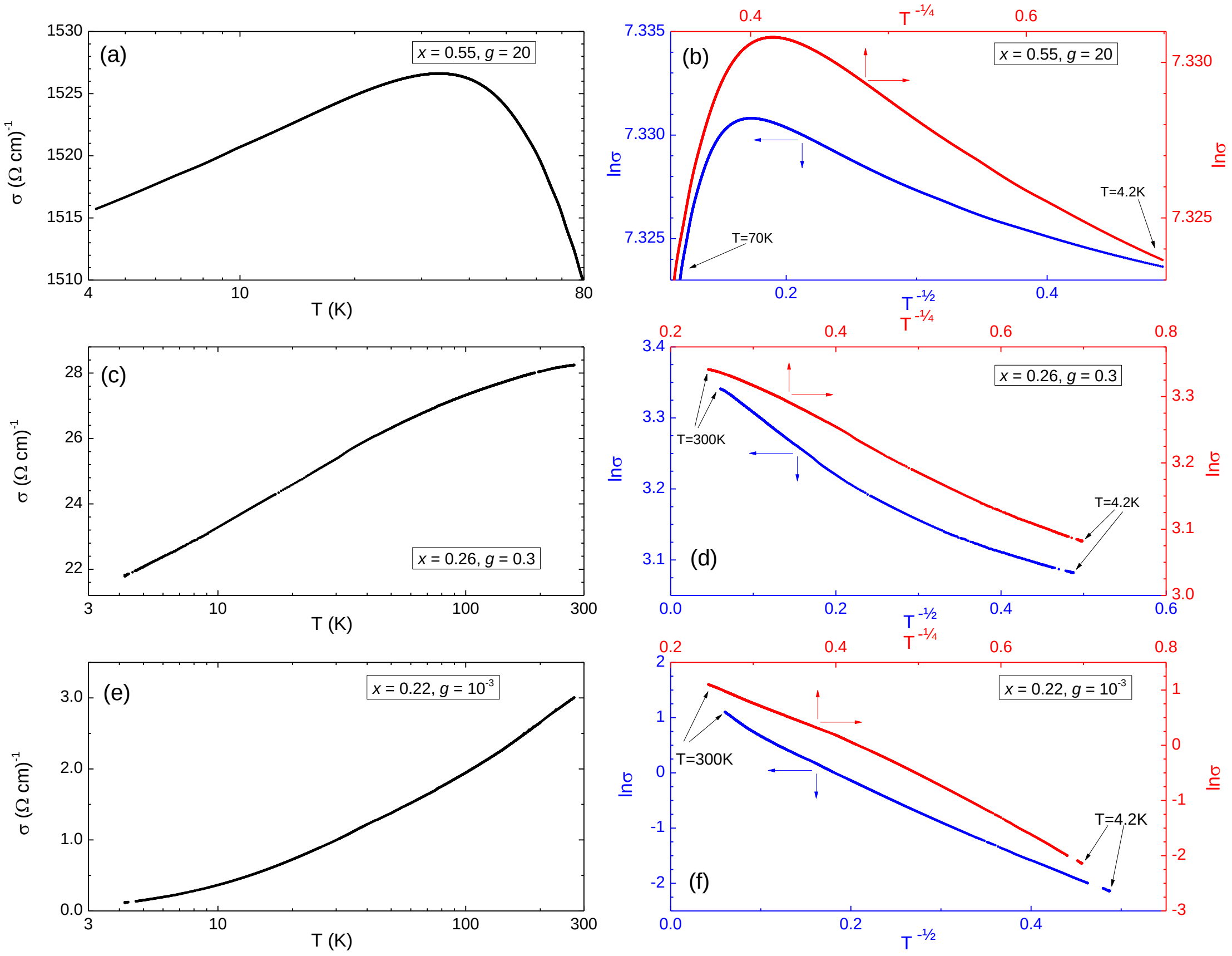

Fig. 4

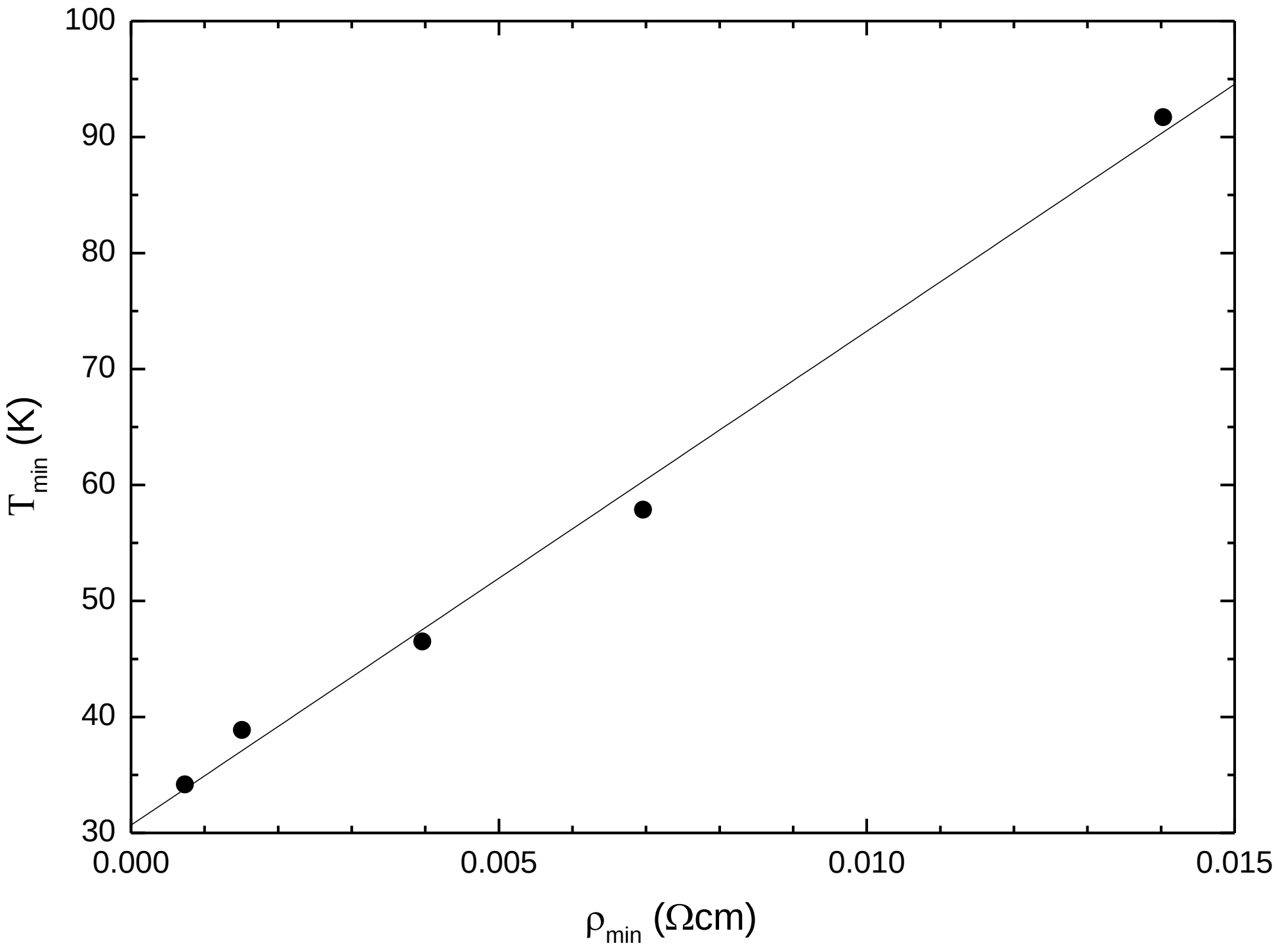

Fig. 5

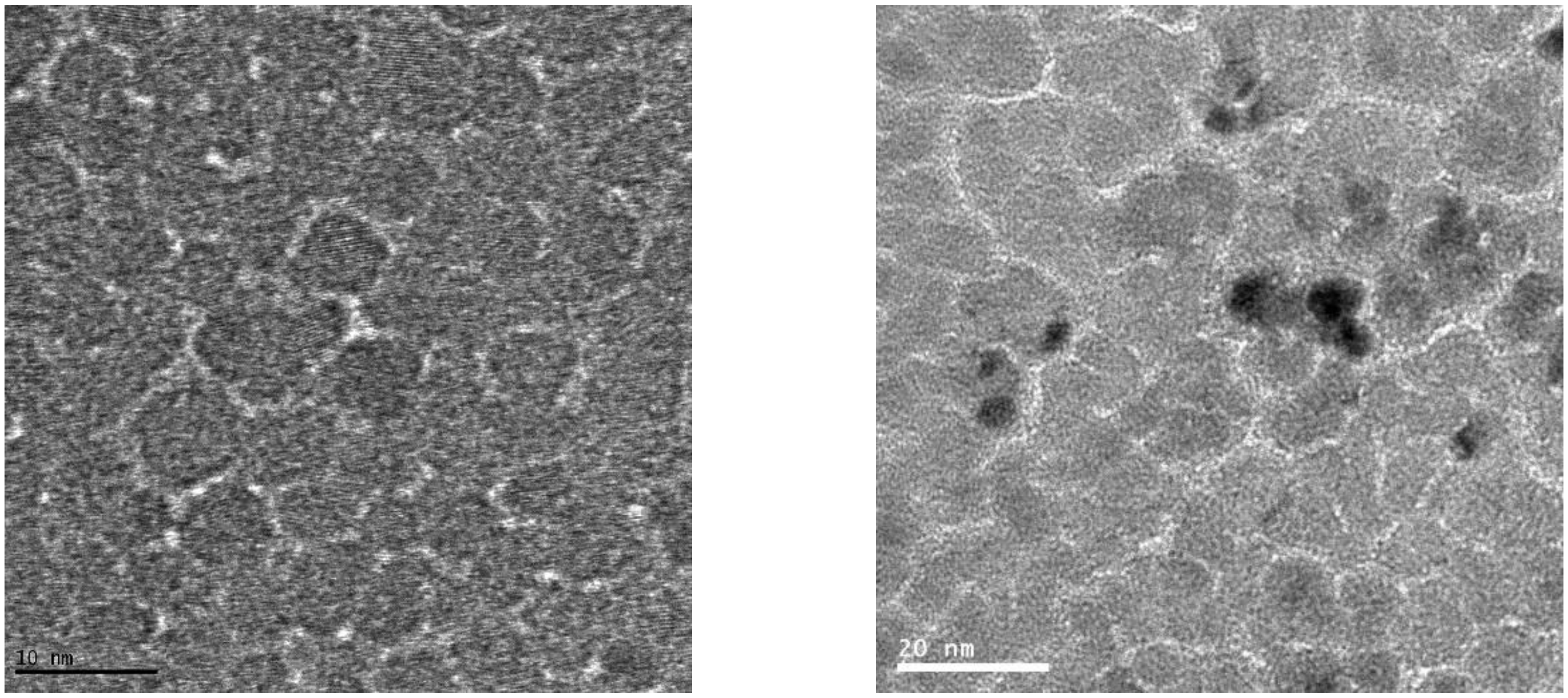

Fig.6

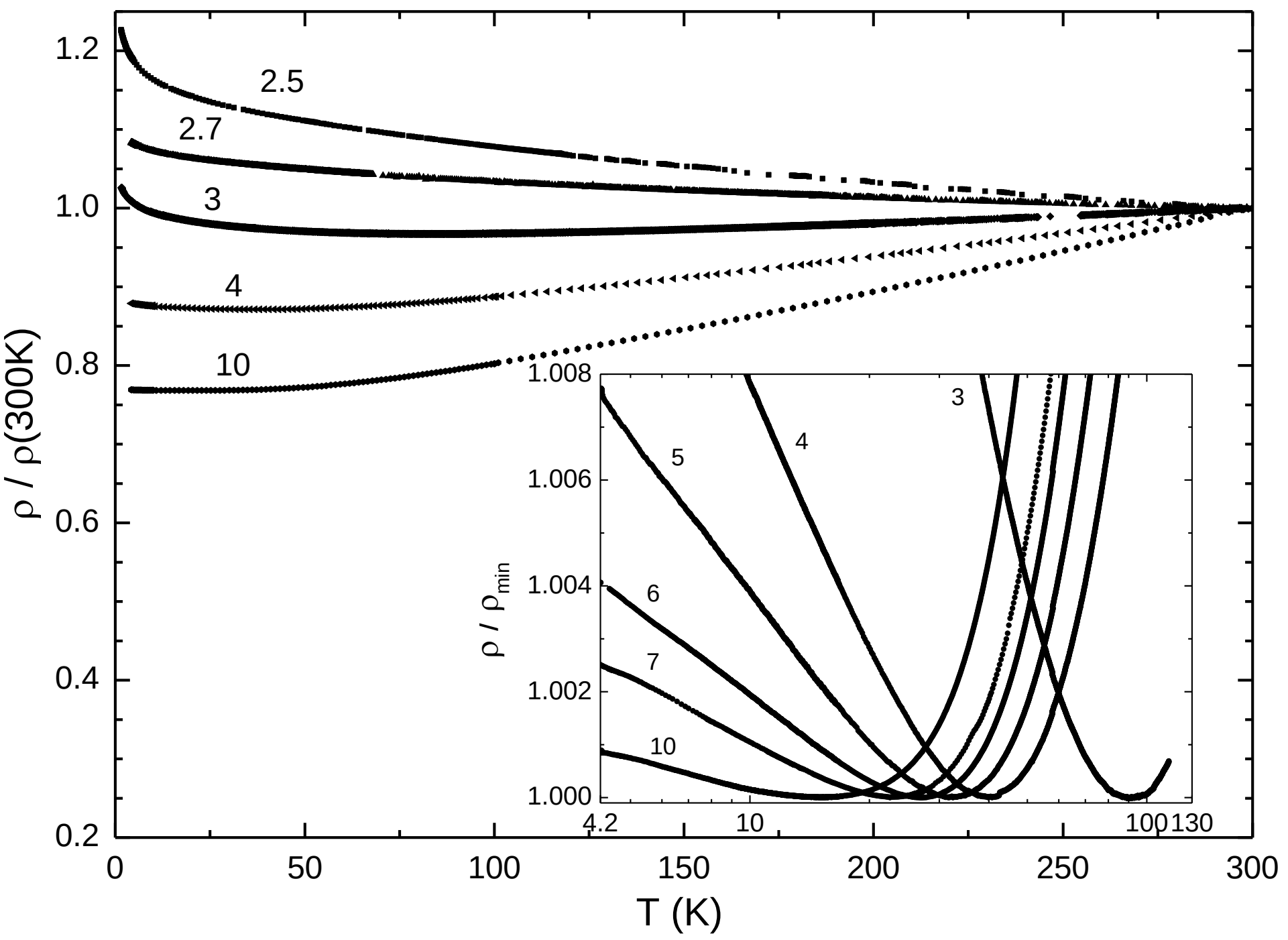

Fig. 7

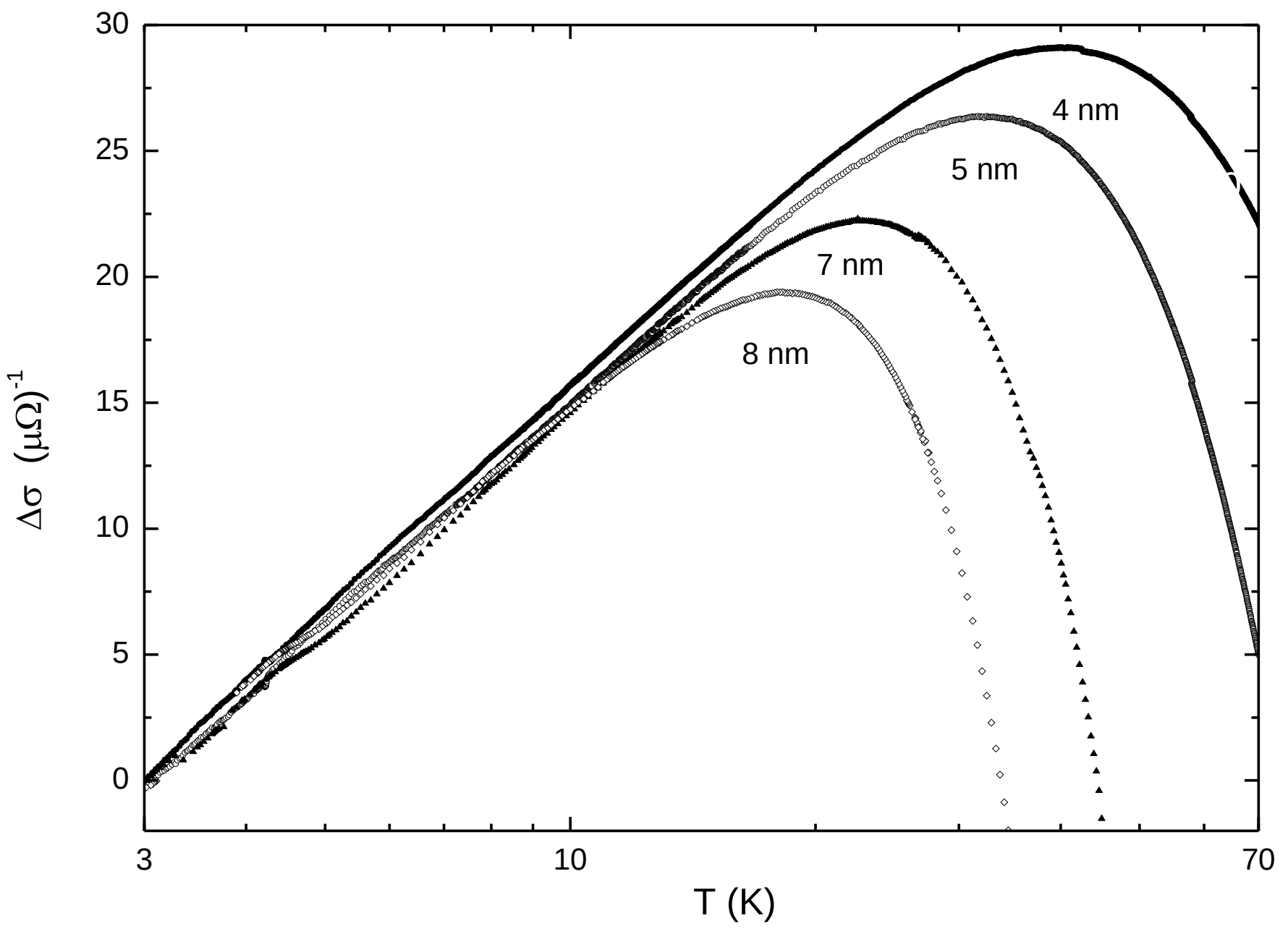

Fig. 8a

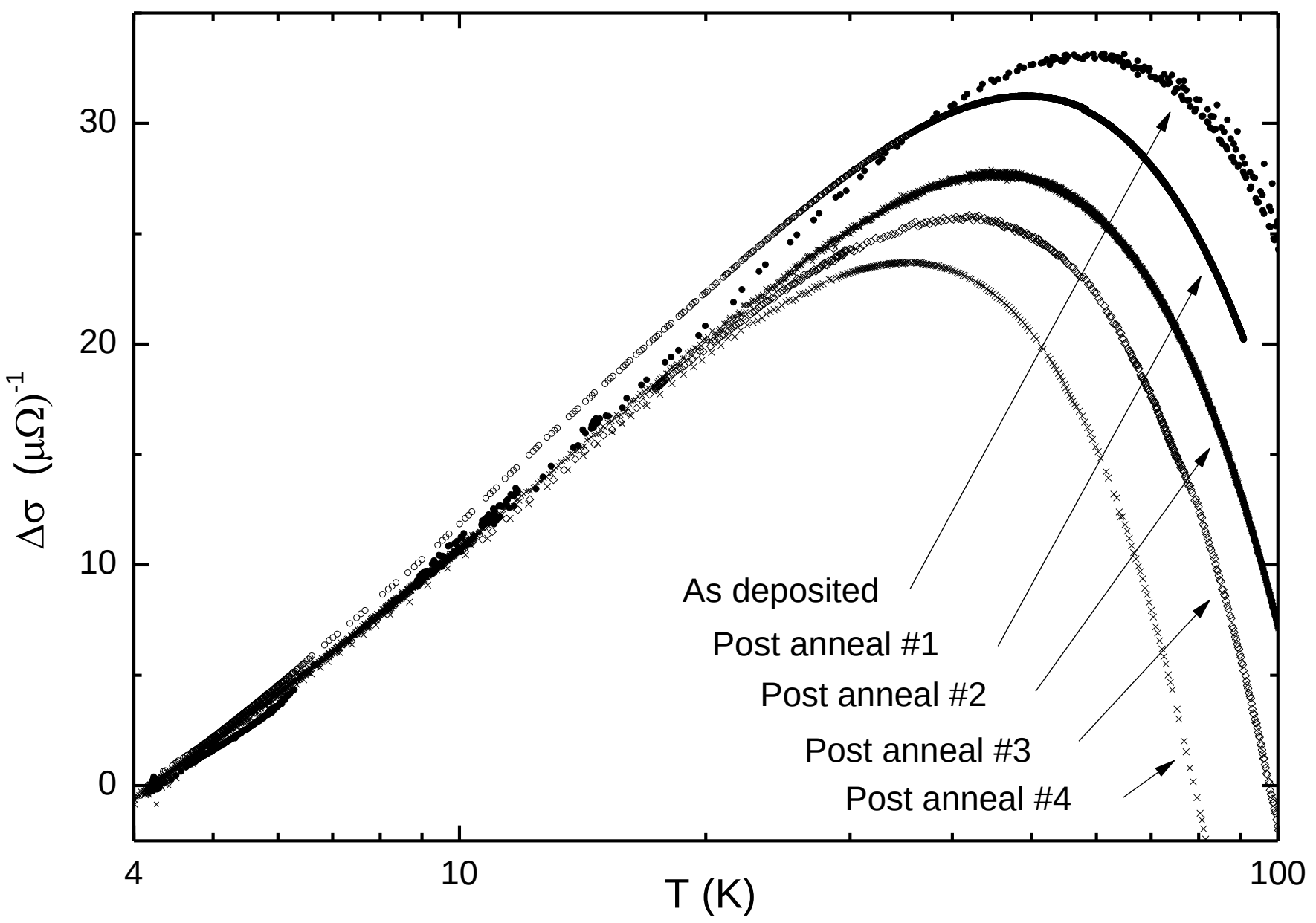

Fig. 8b

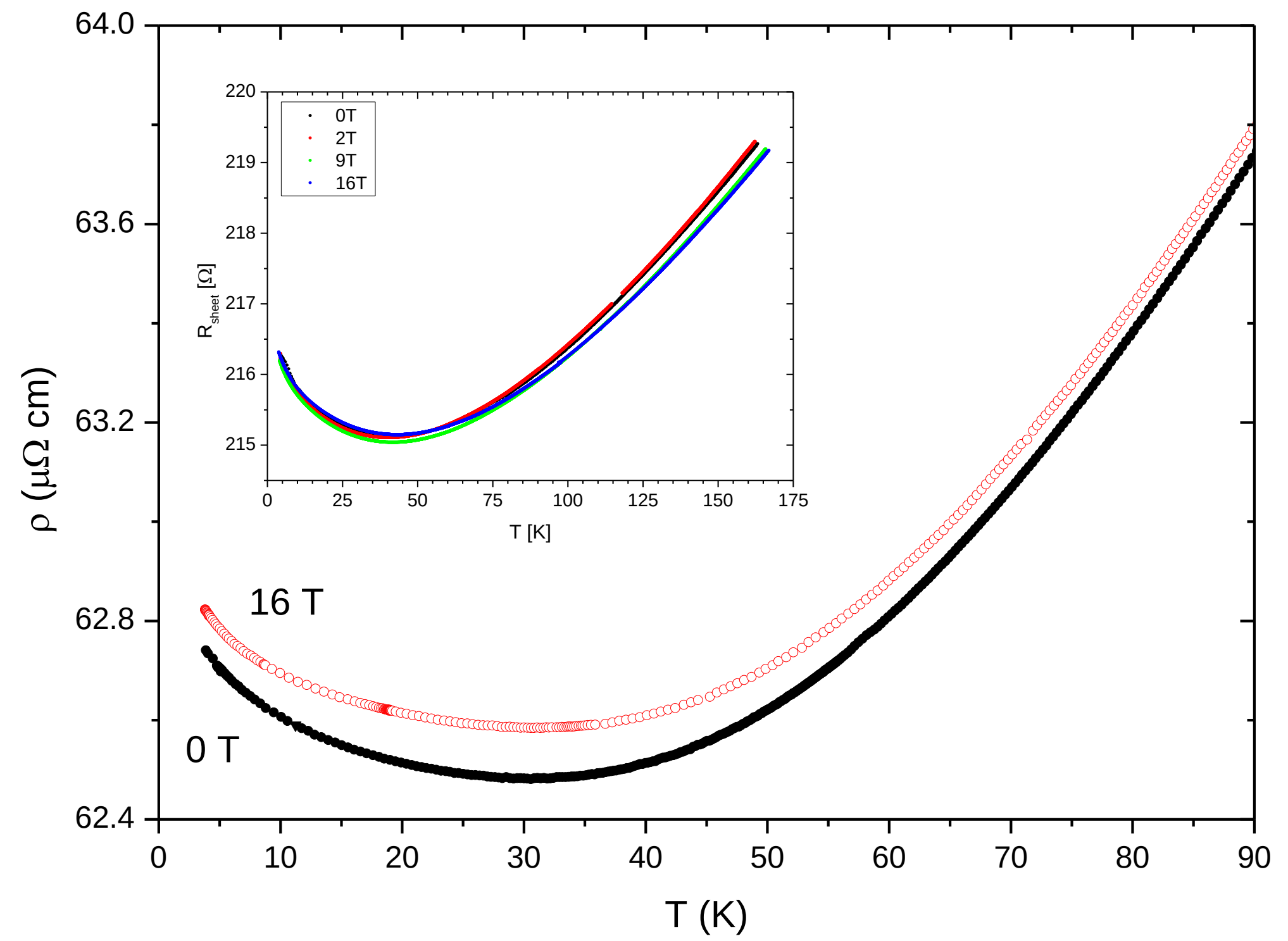

Fig. 9

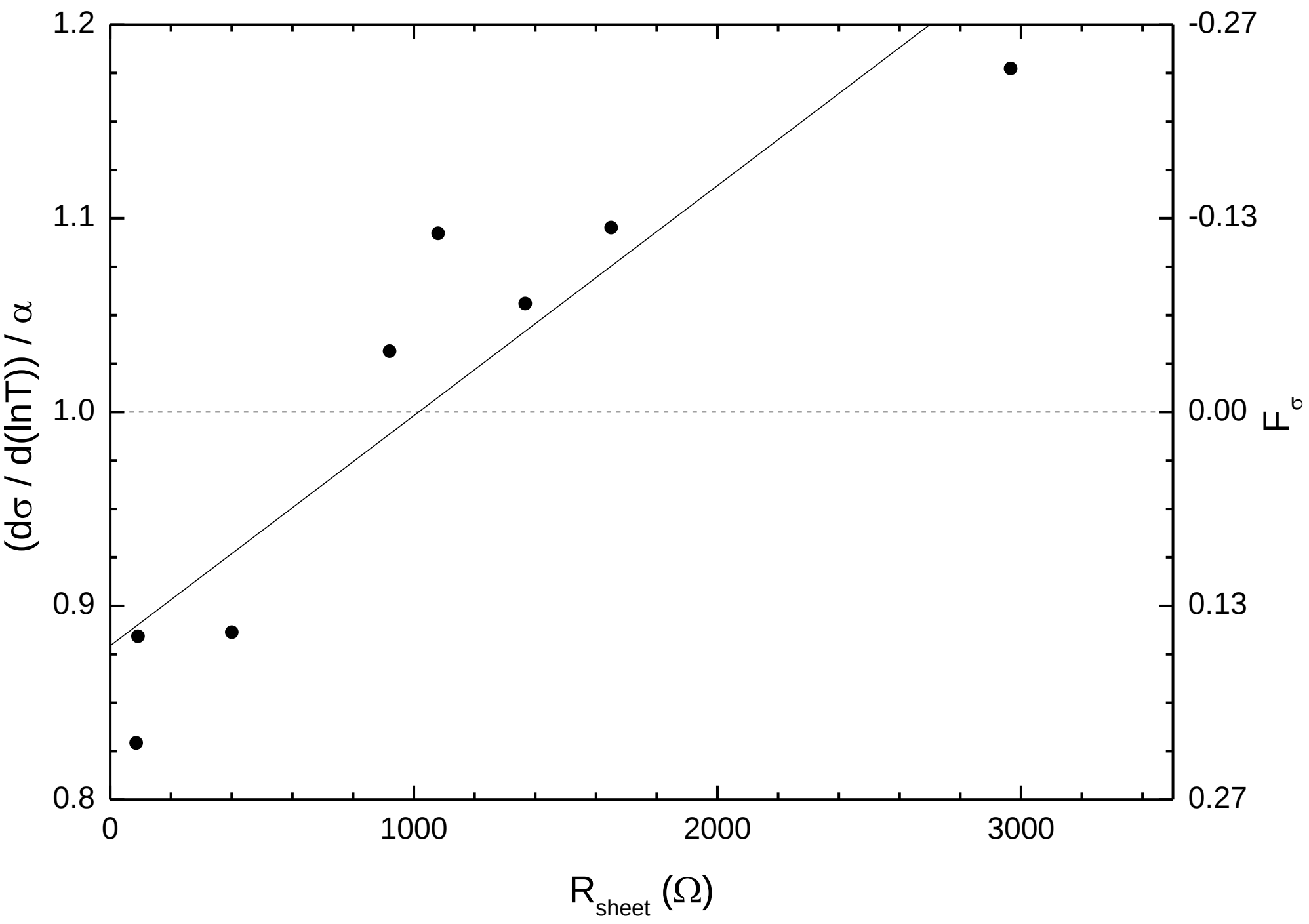

Fig. 10

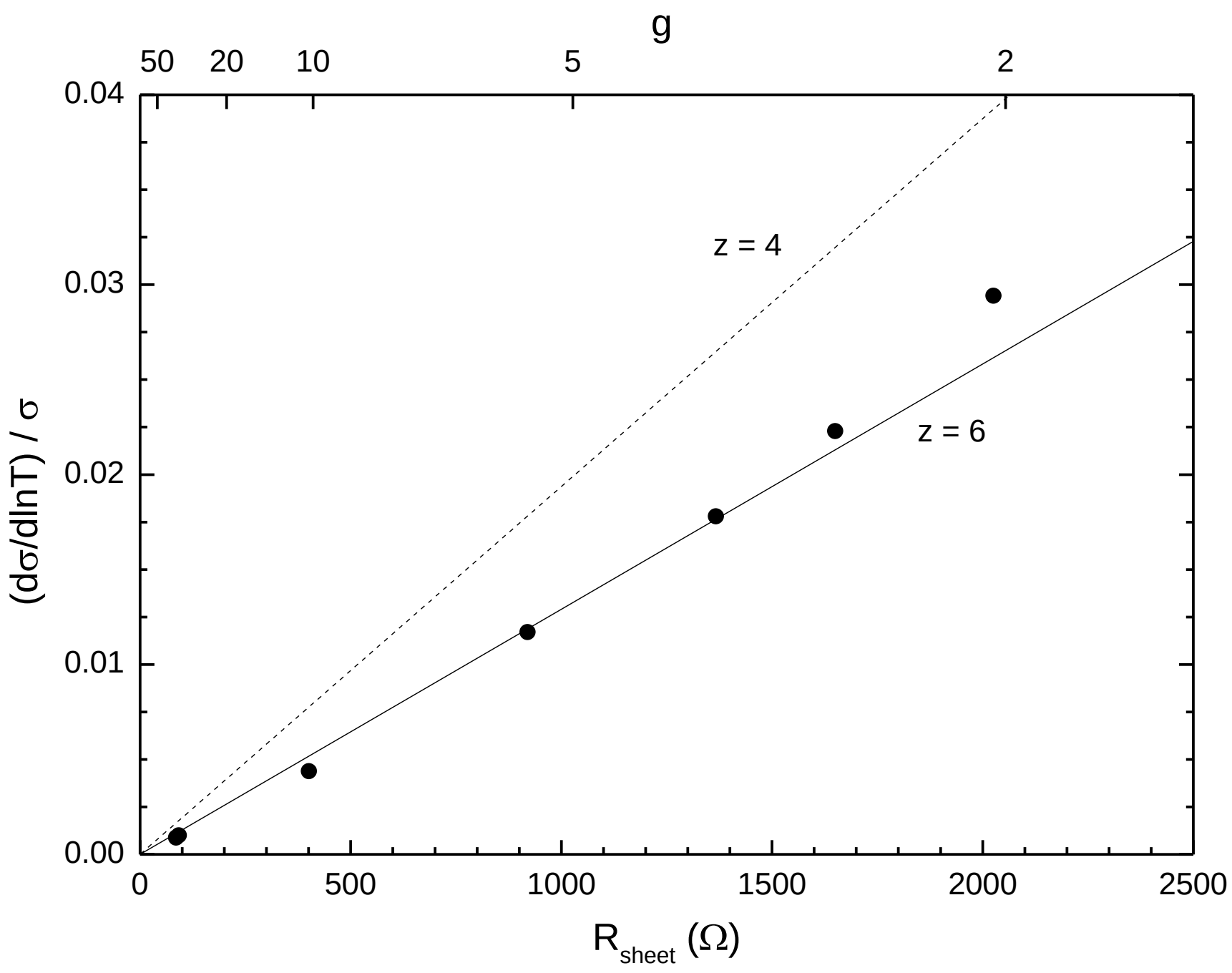

Fig. 11a

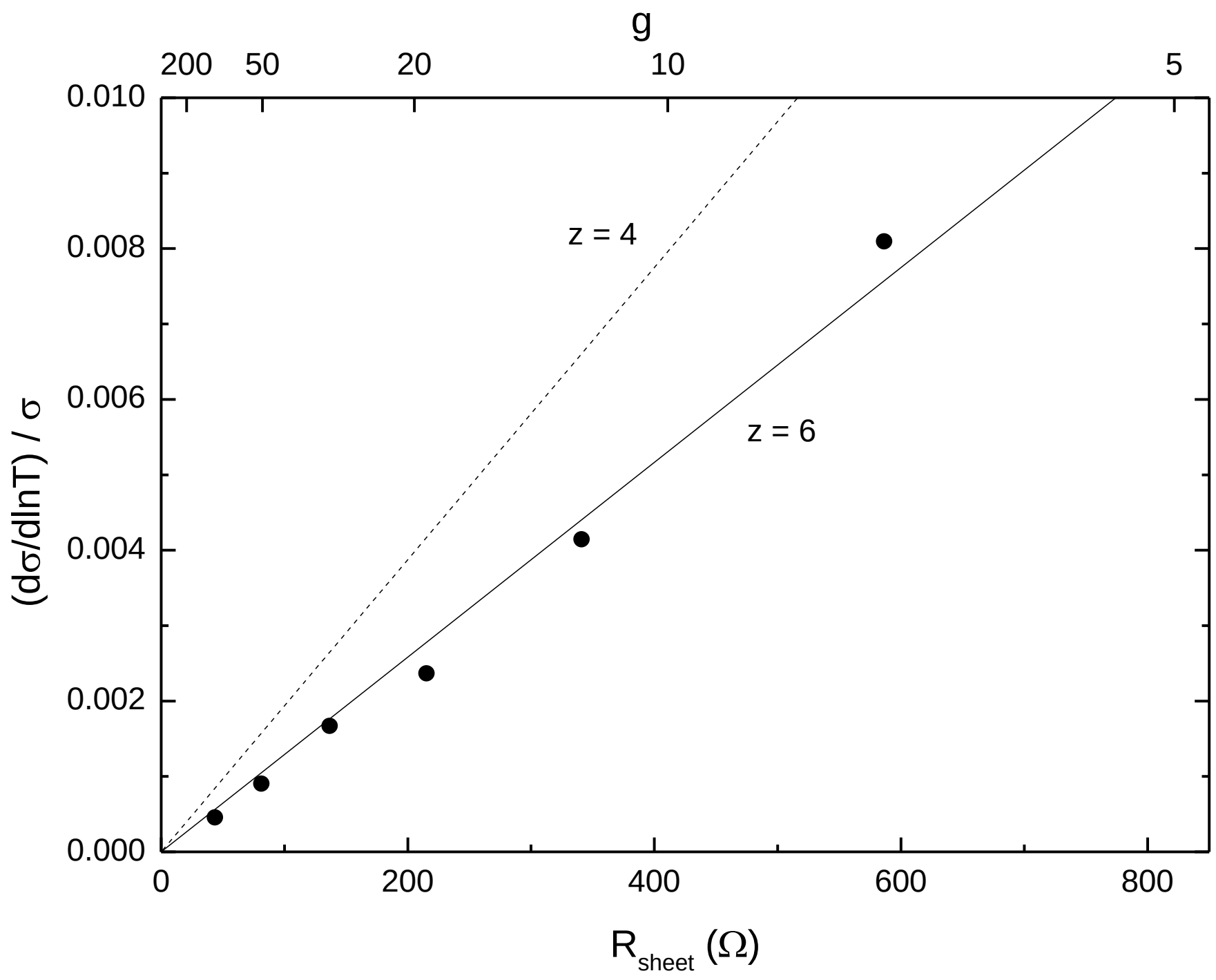

Fig. 11b

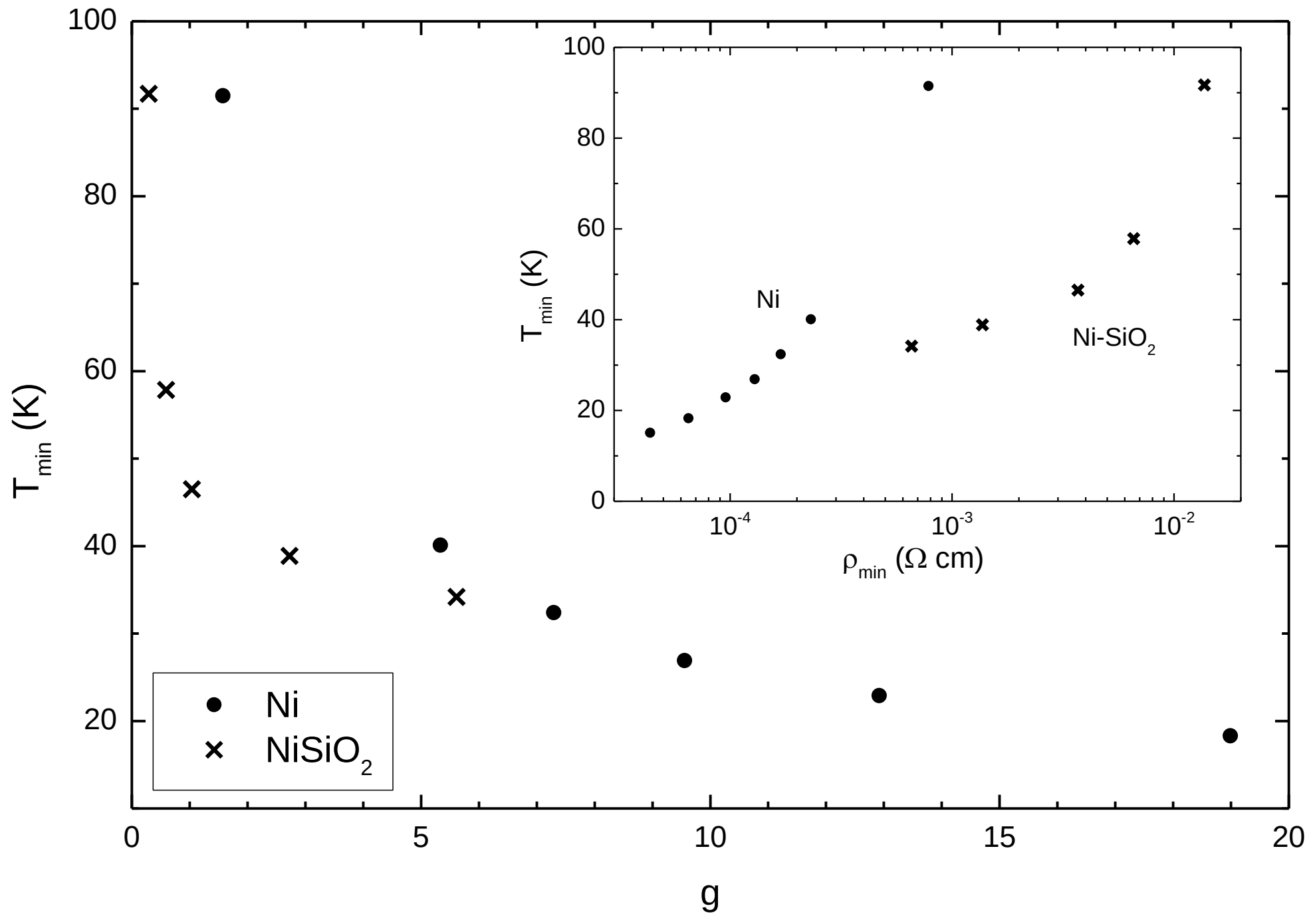

Fig. 12